\documentclass{iau} 
\usepackage{graphicx}

\title[Analytical disk solution in comparison with simulations] %% give here short title %%
{Analytical solution for magnetized thin accretion disk in comparison
with numerical simulations}

\author[Miljenko \v{C}emelji\'{c}, Varadarajan Parthasarathy, W\l odek Klu\'{z}niak] %% give here short author list %%
{Miljenko \v{C}emelji\'{c},
  Varadarajan Parthasarathy
\and W\l odek Klu\'{z}niak}

\affiliation{Nicolaus Copernicus Astronomical Center, Bartycka 18, 00-716
Warsaw, Poland
 \\ email: {\tt miki,varada,wlodek@camk.edu.pl} \\[\affilskip]
}
\pubyear{2018}
\volume{346}  %% insert here IAU Symposium No.
\jname{High-mass X-ray binaries: illuminating the passage from
massive binaries to merging compact objects}
\editors{A.C. Editor, B.D. Editor \& C.E. Editor, eds.}
\begin{document}

\maketitle

\begin{abstract}
We obtained equations for a thin magnetic accretion disk, using the method
of asymptotic approximation. They cannot be solved analytically-without
solutions for a magnetic field in the magnetosphere between the star and the
disk, only a set of general conditions on the solutions can be derived.
To compare the analytical results with numerical solutions, we find
expressions for physical quantities in the disk, using our results from
resistive and viscous star-disk magnetospheric interaction simulations. 

\end{abstract}

\firstsection % if your document starts with a section,
\section{Introduction}
Gravitational infall of matter onto a rotating central object naturally forms
a rotating accretion disk. Matter from the disk is then fed inwards through
an accretion column. Examples of single objects with a disk around them are
protostars and young stellar objects, and in close binary systems a disk
can form when matter from donor star falls onto a white dwarf or a neutron
star.

Analytical hydro-dynamical model of a thin accretion disk, with viscosity
parameterized by Shakura \& Sunyaev $\alpha$-prescription, has been given in
\cite[Klu\'{z}niak \& Kita (2000)]{KK00}. We extend this model, obtaining
the equations for a magnetic thin disk. Analytical solution in the magnetic
case cannot be given without knowing the solutions in a star-disk
magnetosphere, only general conditions on solutions could be derived.

We perform numerical simulations of star-disk magnetospheric interaction,
adding a stellar rotating surface and a magnetic field to the analytical
hydro-dynamical disk solution used as an initial condition in simulations.
A quasi-stationary solution is obtained, from which we find simple matching
expressions for physical quantities in the disk. Those expressions can be
compared with requirements from the analytically obtained equations for the
magnetic disk and with the hydro-dynamic analytical and numerical solutions.

%(Fig.\,\ref{fig1})
\begin{figure}%[t]
% \vspace*{-2.0 cm}
\begin{center}
 \includegraphics[width=5in]{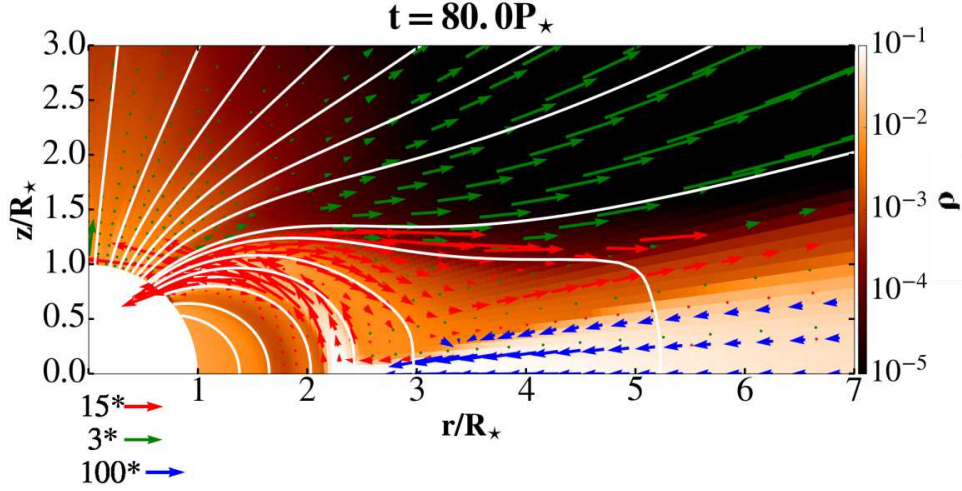} 
% \vspace*{-1.0 cm}
 \caption{
A zoom into our simulation result after T=80 rotations of the underlying star.
Shown is the density in logarithmic color grading in code units, with a sample
of magnetic field lines, depicted with white solid lines. Velocities in the
disk, column and stellar wind are shown with vectors, depicted in different
normalizations with respect to the Keplerian velocity at the stellar surface.}
\label{fig1}
\end{center}
\end{figure}

\section{Numerical setup}
We use the publicly available {\sc pluto} code (v.4.1) (\cite[Mignone
\etal\ 2007; 2012]{m07,m12}), with logarithmically stretched grid in
radial direction in spherical coordinates, and uniformly spaced latitudinal
grid. Resolution is $R\times\theta$=[$217\times100$] grid cells, stretching
the domain to 30 stellar radii. We perform axisymmetric 2D star-disk
simulations in resistive and viscous magneto-hydrodynamics, following
\cite[Zanni \& Ferreira (2009)]{zf09} - see also \cite[\v{C}emelji\'{c}
\etal\ (2017)]{cpk17}.

The equations solved by the {\sc pluto} code are, in the cgs units:
\begin{eqnarray}
\frac{\partial\rho}{\partial t}+\nabla\cdot(\rho\mathbf{v}) =0 \nonumber \\
\frac{\partial\rho\mathbf{v}}{\partial t}+\nabla\cdot\left[\rho\mathbf{v}
\mathbf{v}+\left(P+\frac{\mathbf{B}\mathbf{B}}{8\pi}\right)\mathbf{I}-
\frac{\mathbf{B}\mathbf{B}}{4\pi}-\mathbf{\tau}\right]=\rho\mathbf{g} \nonumber \\
\frac{\partial E}{\partial t}+\nabla\cdot\left[\left(E+P+\frac{\mathbf{B}
\mathbf{B}}{8\pi}\right)\mathbf{v}+
\underbrace{\eta_{\rm m}\mathbf{J}\times \mathbf{B}/4\pi - \mathbf{v}\cdot
\mathbf{\tau}}_{\rm heating\ terms}\right]=\rho\mathbf{g}\cdot\mathbf{v}
-\underbrace{{\Lambda}}_{\rm cooling} \nonumber \\
\frac{\partial\mathbf{B}}{\partial t}+\nabla\times(\mathbf{B}\times\mathbf{v}+\eta_{\rm m}\mathbf{J})=0. \nonumber
\label{eqsmot}
\end{eqnarray}
Symbols have their usual meaning. The underbraced Ohmic and viscous
heating terms and the cooling term are removed in our computations,
to prevent the thermal thickening of the accretion disk. This equals
the assumption that all the heating is radiated away from the disk.
The solutions are still in a non-ideal magneto-hydrodynamics regime,
because of the viscous term in the momentum equation, and the
resistive term in the induction equation.

\section{Analytical solutions ver. numerical solutions}
\begin{figure}[t]
% \vspace*{-2.0 cm}
\begin{center}
\includegraphics[width=2.5in]{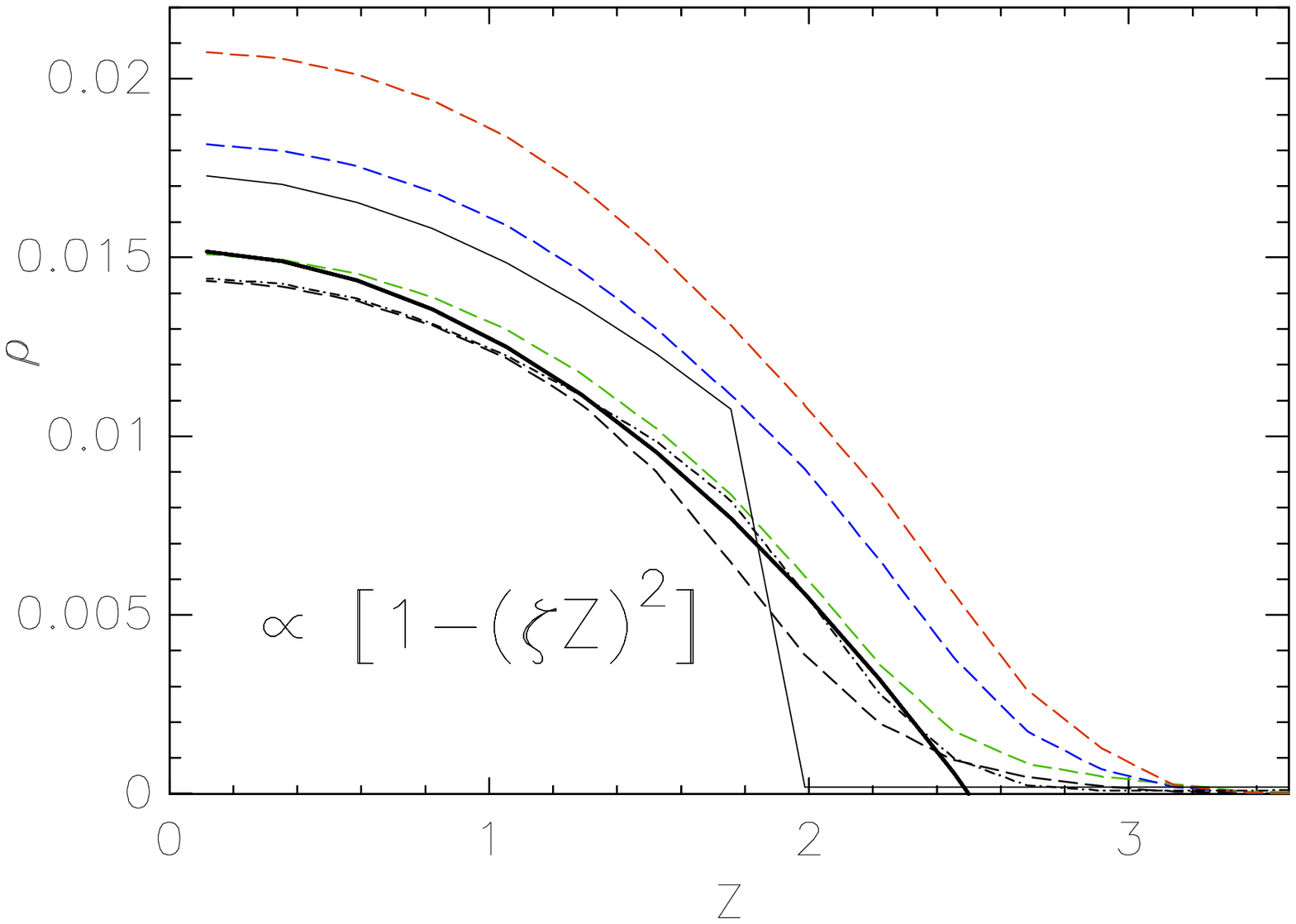}
\includegraphics[width=2.6in]{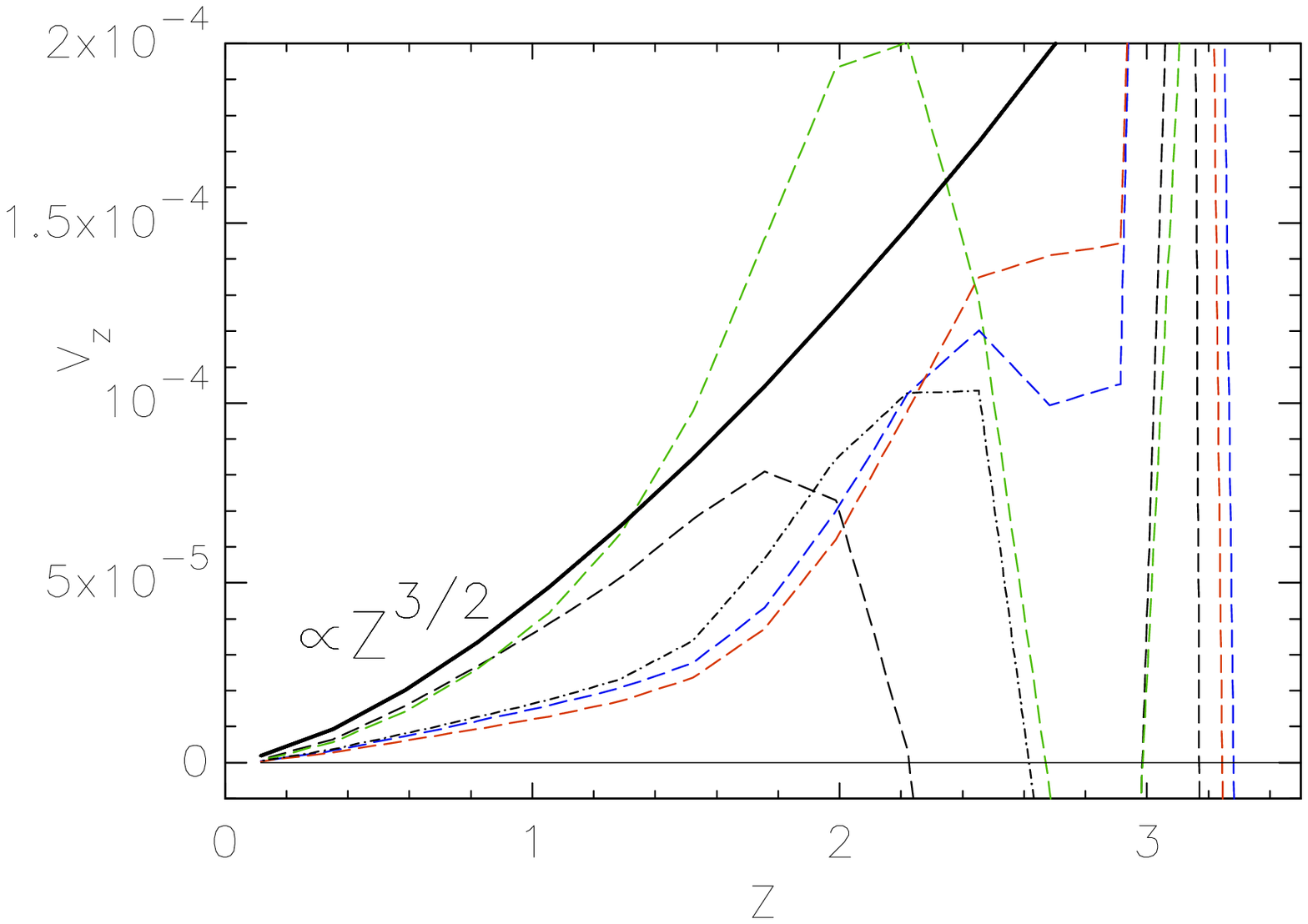} 
\includegraphics[width=2.6in]{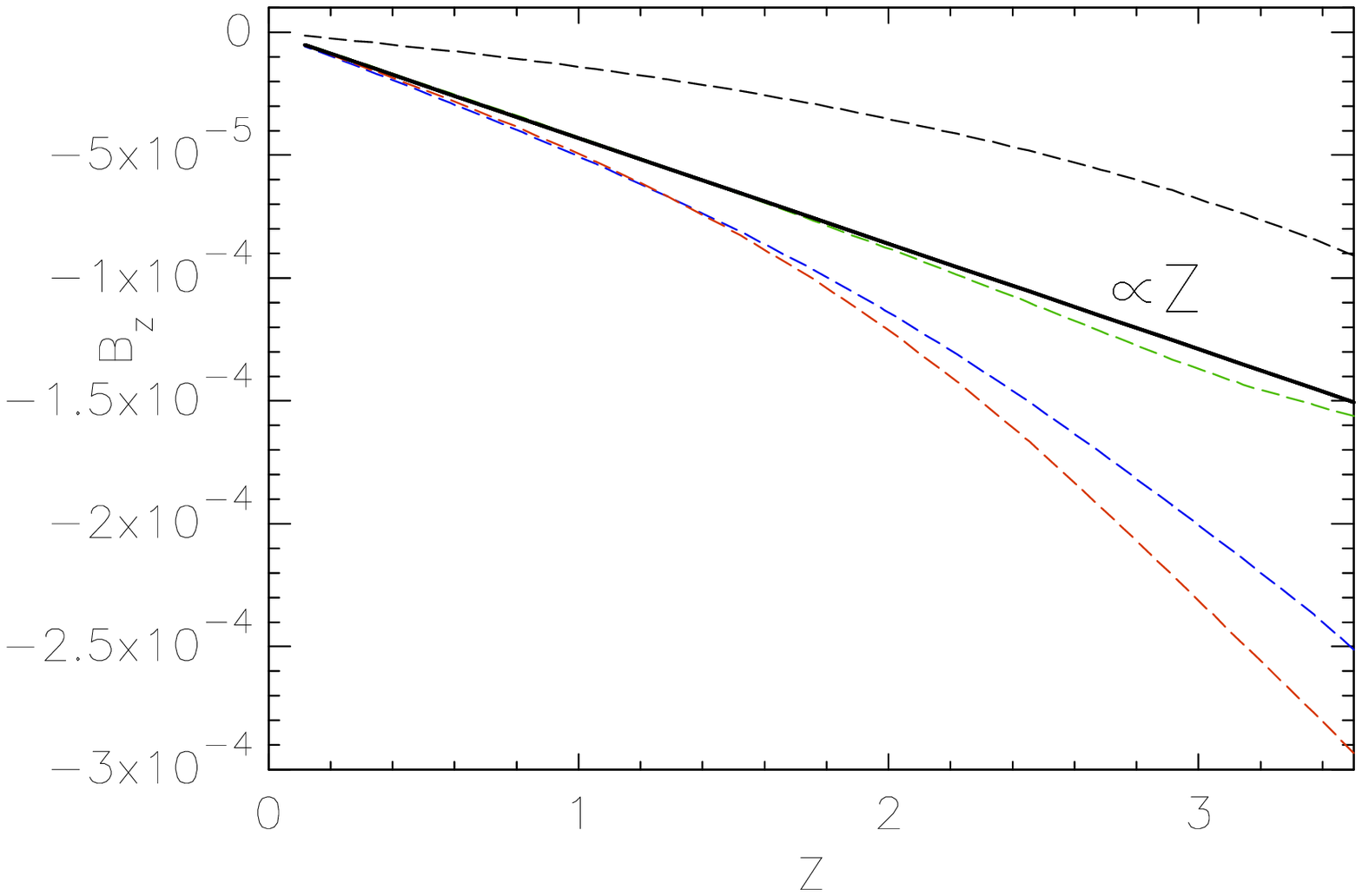}
\includegraphics[width=2.55in]{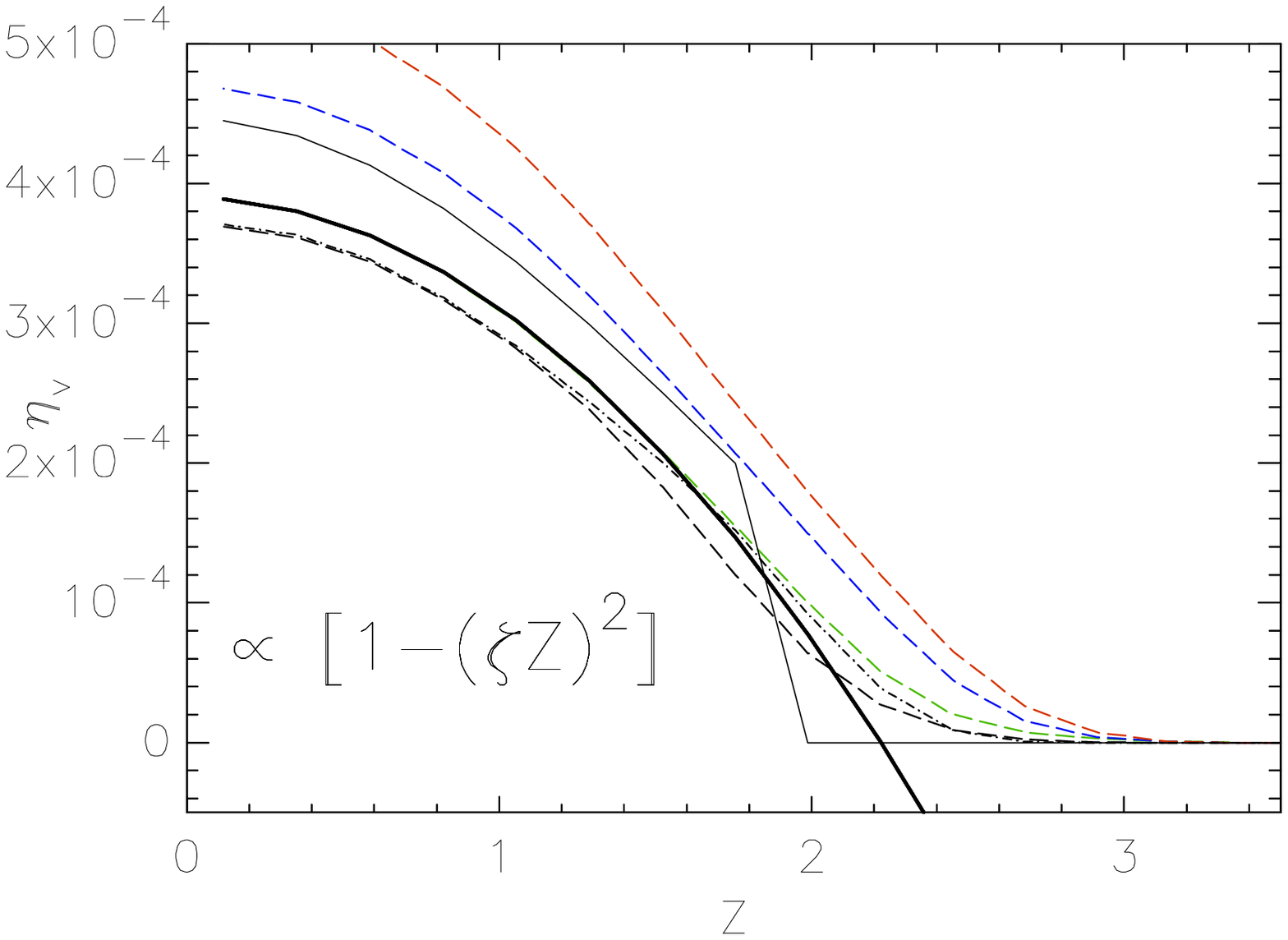}
 \caption{
Trends with the increasing stellar magnetic strength in the disk density,
velocity, magnetic field and viscosity. The values are taken  at R=15R$_\star$
in the vertical direction, with Z expressed in the units of R$_\star$. With
the thin solid line is shown the KK00 solution, with the dash-dotted line is shown
the solution without the magnetic field, and in black, green, blue and green
dashed lines are shown the solutions with $B_\star$=0.025, 0.05, 0.075 and
0.1~T, respectively. With the thick solid line is shown a match to the
$B_\star$=0.05~T case. Expressions for the match in B$_\star$=0.05~T case
are written in each panel, with $\zeta$ in the density equal to 0.4, and
in the viscosity to 0.45. The density, velocity, dynamic viscosity and magnetic
field are expressed in the code units
$\rho_{\rm d0}$=$\dot{M}_0/(R_\star^2v_{\rm K\star})$, $v_0=v_{\rm K\star}$,
$\eta_{\rm v0}=\rho_{\rm d0}R_\star v_{\rm K\star}$, and
$B_0=v_{\rm K\star}\sqrt{\rho_{\rm d0}}$, where $\dot{M}_0$ is the free
parameter in the simulations, denoting the disk accretion rate, and $v_{\rm K\star}$
is the Keplerian velocity at the stellar surface.
}\label{fig2}
\end{center}
\end{figure}
\begin{table}
\caption{Coefficients k in the expressions from results in our simulation with
$B_\star$=0.05~T.}
\label{tabsols} 
\centering                          % used for centering table
\begin{tabular}{ c c c c c c c c c c   }        % centered columns
\hline    % inserts single horizontal line
$k_1$ & $k_2$ & $k_3$ & $k_4$ & $k_5$ & $k_6$ & $k_7$ & $k_8$ & $k_9$ \\
0.88 & -0.09 & 3.8$\times10^{-5}$ & 0.255 & -0.4 & -0.15 & -1.11 & 0.006 & 0.01 \\
\hline
\end{tabular}
\end{table}

We extended the asymptotic approximation from a hydro-dynamic
\cite[Klu\'{z}niak \& Kita (2000)]{KK00} thin accretion disk solution
to a magnetic case. Without knowing the solution for a magnetic
field in a star-disk magnetosphere, which is connected with the
solution in the disk, obtained equations cannot be solved analytically.
In addition to the solutions which remain the same as in the hydro-dynamical
case, only general constraints on the magnetic solution could be derived.

We compare the obtained analytical solutions and constraints in the magnetic
case with the results from simulations. To do this, we write the solutions
from simulations in terms of matching expressions-which are not
formal fits, but simple matching functions chosen to represent the result within
10\% of the value from the simulation, when there is no oscillations. In all
the cases, the match is chosen to the solution in the middle of the
disk, further away from the star than the corotation radius, where the disk co-rotates
with the star, which is at $R_{\rm cor}=2.9R_\star$.

We confirm that in the middle part of the disk, at R=15R$_\star$, the
numerical solution in the magnetic case does not differ much from the
hydro-dynamical analytical and numerical solutions. The difference is only
in the proportionality coefficients $k$ in the expressions for physical
quantities and in the corresponding power laws.

The expressions we obtain are:
\begin{eqnarray}
\rho(r,z)=\frac{k_1}{r^{3/2}}[1-(0.4 z)^2],\nonumber \\
v_r(r,z)=\frac{k_2}{r^{3/2}}[1+(0.5 z)^2],\ v_z(r,z)=\frac{k_3}{r}z^{3/2},\ v_{\varphi}(r,z)=\frac{k_4}{\sqrt{r}},\nonumber\\
B_r(r,z)=\frac{k_5}{r^3}z,\ B_z(r,z)=\frac{k_6}{r^3}z,\ B_{\varphi}(r,z)=\frac{k_7}{r^3}z, \nonumber \\
\eta(r,z)=\frac{k_8}{r}[1-(0.45 z)^2],\ \eta_{\rm m}(r,z)=k_9\sqrt{r}[1-(0.3 z)^2].\nonumber 
\label{eqs1}
\end{eqnarray}

We tabulate the coefficients $k$ in the case of a Young Stellar Object
rotating with $\Omega_\star=0.2$ of the breakup angular velocity, with the
stellar field B$_\star$=0.05~T, anomalous viscosity and resistivity coefficients
$\alpha_{\rm v}$=1 and $\alpha_{\rm m}=1$, in Table~\ref{tabsols}.

In Fig.~\ref{fig2} we show the trends in the density, viscosity and vertical
velocity and magnetic field components in the disk, in the cases with
increasing stellar magnetic field strength.

\section{Conclusions}
We present our results in numerical simulations of a star-disk system with
magnetospheric interaction in the case of Young Stellar Object with the
stellar field of 0.05~T, rotating with 20\% of the breakup velocity.
Quasi-stationary solutions in the disk are obtained, and we find simple
expressions to match the physical quantities.

We find that the expressions in the magnetic cases differ from the results
in the hydro-dynamical simulation and in the analytical solution only in
proportionality coefficients.

In future work we will compare the trends in numerical solutions in the
cases with different stellar magnetic field strength, rotation rate,
viscosity and resistivity.

\begin{acknowledgements}
M\v{C} developed the setup for star-disk simulations while in CEA, Saclay,
under the ANR Toupies grant. His work in NCAC Warsaw is funded by a Polish
NCN grant no. 2013/08/A/ST9/00795 and a collaboration with Croatian
STARDUST project through HRZZ grant IP-2014-09-8656 is acknowledged. VP
work is partly funded by a Polish NCN grant 2015/18/E/ST9/00580. We
thank IDRIS (Turing cluster) in Orsay, France, ASIAA/TIARA (PL and XL
clusters) in Taipei, Taiwan and NCAC (PSK cluster) in Warsaw, Poland, for
access to Linux computer clusters used for the high-performance computations.
We thank the {\sc pluto} team for the possibility to use the code.
\end{acknowledgements}

\end{document}